\documentstyle[psfig,11pt]{article}
\def \AN {{\it Astron. Nach.}}
\def \SAIT {{\it Mem. Soc. Astron. It.}}

\def \AAP {{\it Astron. Astrophys.}}
\def \AAL {{\it Astron. Astrophys. Lett.}}

\def \AAS {{\it Astron. Astrophys. Suppl. Ser.}}

\def \ANNREV {{\it Ann. Rev. Astron. Astrophys.}}
\def \APJ {{\it Astrophys. J.}}
\def \APJL {{\it Astrophys. J. Lett.}}
\def \APJS {{\it Astrophys. J. Suppl.}}

\def \ASR {{\it Adv. Space Res.}}

\def \MN {{\it Mon. Not. R. Astr. Soc.}}

\def \NAT {{\it Nature}}

\begin{document}
\vspace{1.0cm}
{\Large \bf The BeppoSAX view of the hard X--ray background}

\vspace{1.0cm}

A. Comastri$^1$, F. Fiore$^{2,3,4}$, P. Giommi$^3$, F. La Franca$^5$, 
M. Elvis$^4$, G. Matt$^5$, S. Molendi$^6$, G.C. Perola$^5$  

\vspace{1.0cm}
$^1${\it Osservatorio Astronomico di Bologna,  
         via Ranzani 1, I--40127 Bologna, Italy }\\
        \centerline{comastri@astbo3.bo.astro.it}\\
$^2${\it Osservatorio Astronomico di Roma,  
          via Frascati 33, I--00044, Monteporzio Catone, Italy}\\
    \centerline{fiore@quasar.mporzio.astro.it} \\
$^3${\it SAX Science Data Center c/o Nuova Telespazio, 
         via Corcolle 19, I--00131, Roma, Italy}\\
    \centerline{giommi@napa.sdc.asi.it} \\
$^4${\it Harvard Smithsonian Center for Astrophysics, 
         60 Garden street, MA--02138, Cambridge, USA}\\
    \centerline{elvis@head-cfa.harvard.edu} \\
$^5${\it Dipartimento di Fisica Universit\`a  ``Roma Tre",  
         via della Vasca Navale 84, I--00146, Roma, Italy}\\
 \centerline{matt@haendel.fis.uniroma3.it, lafranca@amaldi.fis.uniroma3.it, 
perola@amaldi.fis.uniroma3.it} \\
$^6${\it Istituto di Fisica Cosmica e Tecnologie Relative - CNR, 
         via Bassini 15, I--20133 , Milano, Italy}\\ 
     \centerline{silvano@ifctr.mi.cnr.it} \\
\vspace{0.5cm}

\section*{ABSTRACT}

First results on a medium--deep X--ray survey in the ``new" 5--10 keV band 
carried out with the MECS detectors onboard BeppoSAX are presented.
The High Energy Llarge Area Survey (HELLAS) is aimed to directly explore a 
band where the energy density of the X--ray background is more than twice than 
that in the soft (0.5--2.0 keV) band.
The optical identification follow-up of the first ten HELLAS 
hard X--ray sources indicate that Active Galactic Nuclei are the dominant  
population at 5--10 keV fluxes of the order of 10$^{-13}$ erg cm$^{-2}$ s$^{-1}$. 
We discuss the implications of these findings for the AGN synthesis models 
for the XRB.

\section{INTRODUCTION}

The recent X--ray surveys at both soft (0.5--2.0 keV) and hard (2--10 keV) 
energies have provided a major improvements in our knowledge of the nature
of faint X--ray sources and on the origin of the X--ray Background (XRB).
Deep X--ray observations in the Lockman hole 
carried out with the ROSAT PSPC and HRI detectors have resolved 
into discrete sources about 70--80\% of the 0.5--2 keV XRB 
(Hasinger et al. 1998). 
Optical identifications of a complete sample of 50 ROSAT sources at the
0.5--2.0 keV flux limit of $\sim 5.5 \times 10^{-15}$ erg cm$^{-2}$ s$^{-1}$
reveal that a large fraction ($\sim$ 85 \%) of 
them are AGNs (Schmidt et al. 1998). 
At higher energies (above 2 keV) the lack of imaging capabilities 
has been a major problem for several years. For this reason only 
a few hundreds of bright ($S_{\rm 2-10 keV} > 3 \times 10^{-11}$ 
erg cm$^{-2}$ s$^{-1}$)
sources discovered by HEAO1 (Piccinotti et al. 1982; Wood et al. 1984) 
were known before the launch of ASCA.
The first ASCA surveys (Georgantopoulos et al. 1997; Cagnoni, Della Ceca, 
Maccacaro 1998; Ueda et al. 1998) provided a 
dramatic improvement (about a factor 300) in terms of limiting flux. 
As a result a significant fraction  ($\sim$ 30 \%) of the 2--10 keV XRB is 
already resolved into discrete 
sources at a limiting flux of $\sim 10^{-13}$ erg cm$^{-2}$ s$^{-1}$.
Programs to optically identify these sources have already started 
(Boyle et al. 1998) suggesting that also at higher energies AGN 
constitute the dominant population and adding further evidence on the fact 
that the bulk of the XRB is made by the integrated emission of Active 
Galactic Nuclei.

Detailed modelling of the XRB spectral intensities in terms of the integrated
contribution of AGNs has been hampered by the fact that the 2--10 keV 
AGN spectra are much steeper than the XRB spectrum (the ``spectral paradox").
It has been proposed (Setti \& Woltjer 1989) that this problem can 
be solved assuming an important contribution from sources with spectral shapes
flattened by absorption. 
Based on this proposal the 3--100 keV XRB spectral intensity (Fig. 1) and 
source counts in different energy bands (Fig. 2)
can be reproduced by the combined emission of Seyfert galaxies
and quasars (type 1) and obscured (type 2) AGNs with a range of column
densities and luminosities
(Matt \& Fabian 1994; Madau et al. 1994; Comastri et al. 1995).

All these models rely on AGN unified schemes, which require, in their strictest
version (Antonucci 1993), a type 1 nucleus in all AGN,
surrounded by a  geometrically thick torus, blocking the
line of sight to the nucleus (continuum and broad lines) in type 2
objects. The same X--ray luminosity function and cosmic evolution 
usually parameterized as pure luminosity evolution, is 
then assumed for both type 1 and type 2 AGN, which imply the existence 
of a population of highly absorbed high luminosity quasars (type 2 QSOs).
Even if the discovery of a few type 2 QSO candidates has been reported
(e.g. Otha et al. 1996) this population has proved to be elusive 
and no compelling evidence for any type 2 QSO exists (Halpern et al. 1998).  
Alternatively, the bulk of the
hard XRB could be made by a larger population of lower luminosity
Seyfert 2 like galaxies, possibly subject to a different 
cosmological evolution,  as predicted
in scenarios where the absorption takes place in a starburst region
surrounding the nucleus (Fabian et al. 1998). 
Is is interesting to note that a new estimate of the evolution of the 
soft X--ray  luminosity function from ROSAT data 
(Miyaji, Hasinger, Schmidt 1999a) indicates a more complex behaviour 
which is best parameterized with luminosity dependent density evolution.
Based on these new results Miyaji, Hasinger, Schmidt (1999b) have worked 
out an AGN synthesis model which is able to reproduce the observational 
constraints. 

Given that the contribution of absorbed sources (irrespective of their
nature and cosmological evolution properties) increases with energy (Fig. 1),
a crucial test for the XRB models and in particular on the 
spectral and evolutive properties of obscured AGNs 
would require the comparison of their predictions with the results of optical
identifications of complete samples of X--ray sources possibly selected in the
hard X--ray band.

\begin{figure}
\centerline{\psfig{figure=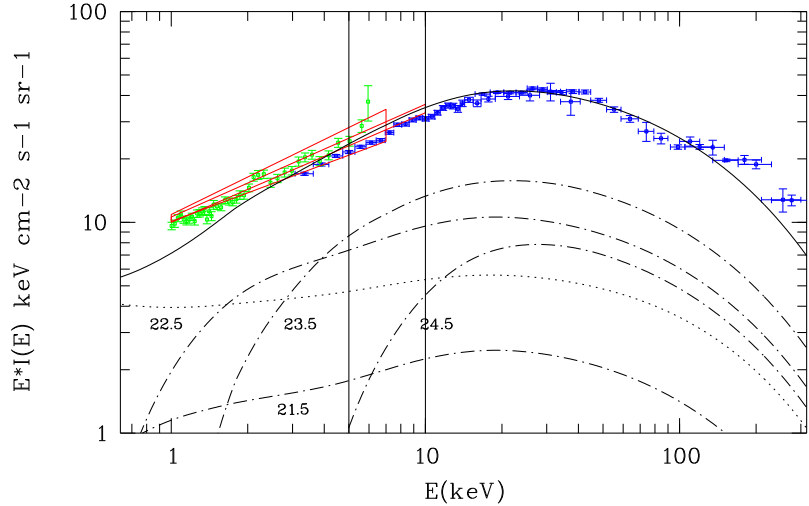,height=8cm,angle=-90}}
\caption{The AGN synthesis model (solid line) fit to the XRB spectrum.
The black points in the 3--300 keV band are from HEAO1, while the grey points
and bow-tie regions in the 1--7 keV band are from ASCA.
The dotted line represents the contribution from unabsorbed objects. 
The dash--dotted lines represent the contribution of obscured AGNs and
are labeled by the logarithm of the column density (see Comastri 1999 for
details). The vertical thick lines
mark the 5--10 keV band of the HELLAS survey. In this band the
energy density of the XRB is about twice that in the soft band.
The increased contribution of absorbed objects is also evident.}
\end{figure}

\begin{figure}
\parbox{8.5truecm}
{\psfig{figure=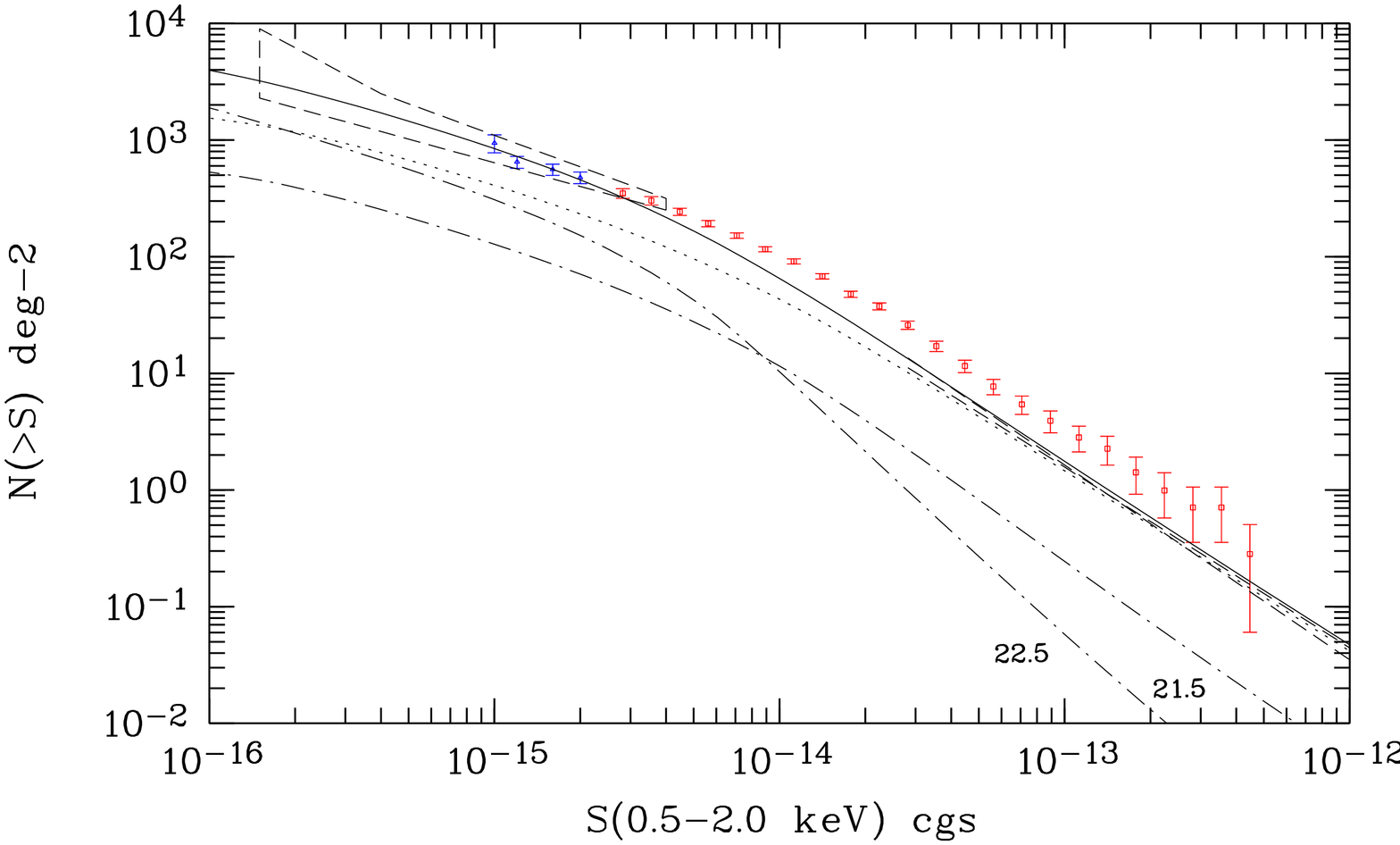,height=8cm,width=8.2cm,angle=-90}}
\  \hspace{0.5truecm}     \
\parbox{8.5truecm}
{\psfig{figure=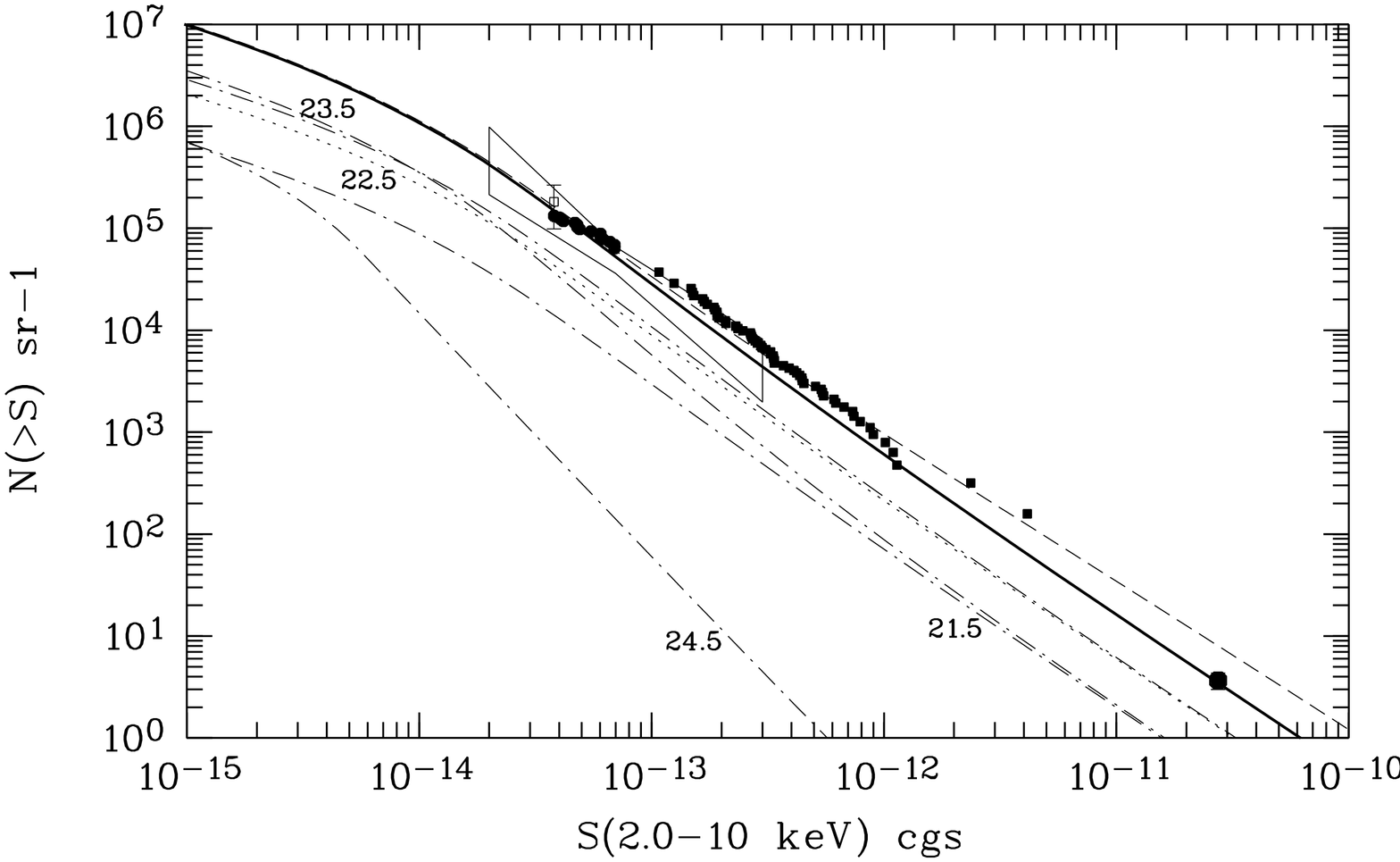,height=8cm,width=8.2cm,angle=-90}}
\caption{The AGN synthesis models predicted counts in the 0.5--2.0 keV
(left panel) and 2--10 keV (right panel) energy ranges compared with 
observational data. The meaning of the various curves is the 
same as in figure 1, the dashed upper curve represents the summed contribution of 
AGNs and clusters of galaxies. 
Data points in the 0.5--2.0 keV band are from Hasinger 
et al. 1998; in the 2--10 keV band, from bright to faint fluxes, are from 
Piccinotti et 
al. 1982; Cagnoni et al. 1998; Giommi et al. 1998; Ogasaka et al. 1998. The 
bow-tie region is from a fluctuation analysis of ASCA data by Gendreau, Barcons
\& Fabian 1998.}
\end{figure}


\section{THE HELLAS SURVEY}

BeppoSAX (Boella et al. 1997a) provides a good opportunity to investigate the hard X-ray
sky, thanks to an improved sensitivity of the MECS
instrument (Boella et al. 1997b above 5 keV (5--10 keV flux limit of
$\sim 0.002$ mCrab in 100 ks, to be compared with the 0.5-1 mCrab flux
limit of the HEAO1-A2 and Ginga surveys), and improved point spread
function (error circles of 1~arcmin, 95\% confidence radius) which
allows optical counterparts to X-ray sources to be identified. 
The BeppoSAX High Energy LLarge Area Survey (HELLAS, Fiore
et al. 1999, in preparation) has cataloged some 170 sources in several 
MECS observations with exposure times between 10 and 300 ks in the 5--10 keV 
band.
Moreover about 500 sources in the 1.5--10 keV band at the flux limit
of $5\times10^{-14}$ erg cm$^{-2}$ s$^{-1}$ have been detected 
in the same fields. The HELLAS sources surface density 
is of the order of 20 sources per square degree
at the limiting flux of $5\times10^{-14}$ erg cm$^{-2}$ s$^{-1}$
(Giommi et al. 1998), implying that between 30 and 40 \% of the XRB 
(depending on its normalization) is already resolved in sources.

In order to investigate the nature of the HELLAS sources 
we have begun a program to identify their optical counterparts.
In particular we want to address the following points:

\begin{itemize}

\item[$\bullet$]
If they are obscured AGNs of the XRB synthesis models
which is the range of absorbing column densities 
and X--ray luminosities ?

\item[$\bullet$]
What is the space density and cosmological evolution
of obscured type 2 AGNs ? Is it different from type 1 ?

\item[$\bullet$]
Are the absorbed sources optically classified as type 2 objects ? 
Is there any relation between X--ray absorption and optical reddening ?

\end{itemize}

\section{THE OPTICAL IDENTIFICATION}

Optical spectroscopy in the 3450--8000 \AA ~ range with a resolution 
of $\sim$ 7 \AA ~ of a subsample of 10 HELLAS sources has been carried
out on March 1998 from the Kitt Peak Observatory 4 meter telescope.
The sources, selected only on the basis of visibility reasons, 
should be representative of the whole HELLAS sample.
Indeed they span a relatively wide range in X--ray fluxes 
($\sim 7 \times 10^{-14}$ -- 10$^{-12}$ erg cm$^{-2}$ s$^{-1}$ in the 5--10 
keV energy range) and X--ray hardness ratios. 
In order to avoid further possible selection biases all the optical 
counterparts, at the magnitude limit of R=20, within each HELLAS error box 
(of about 1 arcmin radius) have been observed. 
The most likely counterpart of the X--ray source has been identified 
among objects with large hard X--ray to optical flux ratio like AGN
Galactic binaries, clusters of galaxies, bright galaxies or stars.
A detailed discussion of the optical and X--ray properties of this 
subsample can be found in Fiore et al. (1999), while the complete
set of optical spectra and finding charts will be presented in
La Franca et al. (1999 in preparation).

Here we briefly summarize the first results of the identification process. 
In eight cases out of ten only one plausible candidate has been
found: three broad line quasars in the redshift range 0.8--1.28
with a blue continuum spectrum; two broad emission line quasars with a very
red optical continuum (0.2 $< z <$ 0.35); 3 narrow emission 
line galaxies identified with type 1.8--2.0 Seyferts on the basis of diagnostic
lines diagrams ($z <$ 0.4). In one error box two optical counterparts 
may contribute to the hard X--ray flux. The diagnostic line ratios suggest
lower excitation than Seyfert galaxies and thus we classify them as LINERS. 
Finally in one case we were not able to find a plausible identification.

The probability of finding 
by chance these sources taking into account the mean surface density
of AGNs and LINERS at the limiting magnitude  R=20 in our error boxes 
is very low ($< 0.2$ \%).
The optical identification follow--up provides thus strong evidence that 
AGNs are the dominant source population in the 5--10 keV band at 
fluxes of the order of $\sim$ 10$^{-13}$ erg cm$^{-2}$ s$^{-1}$.


\section{DISCUSSION}

The X--ray number counts of the HELLAS sources (Giommi et al. 1999 in preparation;
Fiore et al. 1999 in preparation) are compared with the predictions
of the AGN synthesis model of Comastri et al. (1995) in the 5--10 keV band.  
The BeppoSAX error bars (Fig.~3) have been computed assuming a range 
of spectral indices for the count rate to flux conversion and 
the uncertainities in the MECS off--axis sensitivity.
There is a relatively good agreement between data and model predictions at 
the faint flux limit of HELLAS ($\sim$ 5 $\times 10^{-14}$ erg cm$^{-2}$ s$^{-1}$).
The discrepancy at brighter fluxes may be due to the contribution of other 
sources (i.e. stars and galaxies) not included in the model and/or to 
small residual calibration problems. 
A more detailed discussion of the BeppoSAX source counts 
and associated calibration uncertainities in the 5--10 keV band 
will be presented elesewhere (Giommi et al. 1999; Fiore et al. 1999)

\begin{figure}
\centerline{\psfig{figure=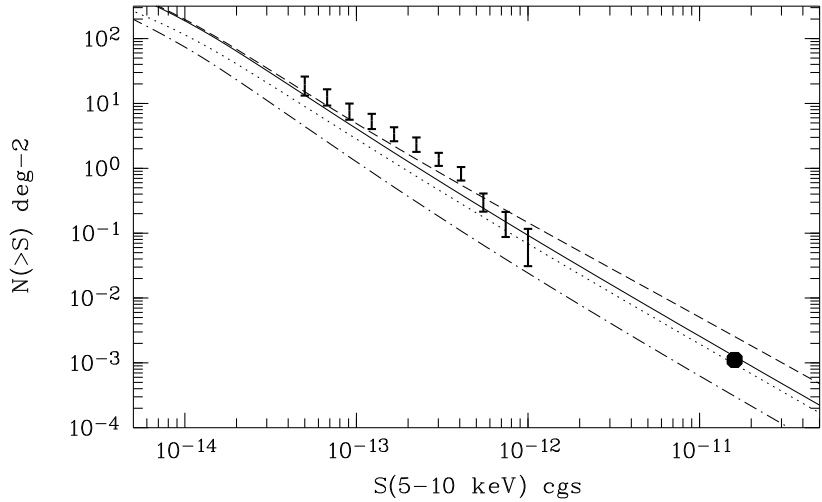,height=8cm,angle=-90}}
\caption{The AGN synthesis model prediction (solid line) 
are compared with the HELLAS log N -- log S error bars 
and at bright fluxes with the HEAO1--A2 AGN counts (Piccinotti et al. 1982) 
converted to the 5--10 keV band assuming a power law spectrum with
energy index $\alpha = 0.8$. The dotted line represents the contribution from 
unabsorbed and relatively unobscured objects (log $N_H <$ 23). 
The dot--dashed line represents the contribution of obscured AGNs
(log $N_H >$ 23). The upper dashed line is a preliminary estimate of the 
summed contribution of AGNs and clusters of galaxies}
\end{figure}

Even if the sample is small we have tried to compare the X--ray and optical
properties of the identified sources. An estimate of the absorbing column
density has been made using the observed hardness ratios and measured redshifts
assuming an average power law spectrum ($F_{\nu} \propto \nu^{-\alpha}$)
with energy index $\alpha$ = 0.8.
We have found evidence for absorption in five sources with estimated
column densities in the range log $N_H$ $\sim$ 22.5--23.2. 
Moreover 4 of the absorbed 
sources show evidence of extinction also in the optical spectrum
(the 3 Seyfert 1.8-2.0 galaxies and 1 red quasar).
The fraction of obscured AGNs seems to be higher than in previous ROSAT and
ROSAT/ASCA surveys (Schmidt et al. 1998; Akiyama et al. 1998) supporting
the AGN synthesis models for the hard XRB. 
In particular the expected number of heavily obscured sources 
(log $N_H >$ 23) according to the Comastri et al. (1995) model is of 
the order of 30\% at the HELLAS
flux limit (dashed line in figure 3) in agreement with the present 
findings.
Clearly a much improved source statistic and further optical identifications 
are needed to confirm these results.


A major improvement in our understanding of the origin of the hard 
X--ray background and on the AGN content of X--ray surveys must await 
the identification of a few hundred hard X--ray selected sources.
The complete identification of the entire HELLAS sample, which is actually
in progress, will allow us to measure the relative fraction of type 2
and type 1 hard X--ray selected AGNs and to study for the first time
the X--ray luminosity function and evolution of obscured AGNs.

This would constitute a crucial piece of information
for the modelling of the XRB. Indeed a significant intrinsic 
dispersion in the X--ray spectral slopes distribution
(as in the case of a significant number of absorbed sources with flat spectra)
can affect the space density and evolution of the entire
population if not properly taken into account. 
In particular as shown by Francis (1993) and Page et al. (1996) 
such a dispersion in the X--ray spectra introduces spurious density 
and/or luminosity evolution and, as a consequence, an incorrect estimate
of the AGN contribution to the XRB.

\section{Acknowledgements}

We thank the BeppoSAX SDC, SOC and OCC team for the successful
operation of the satellite and preliminary data reduction and
screening.  
This research has made use of linearized and
cleaned event files produced at the BeppoSAX Science Data Center.
AC acknowledges partial financial support from the Italian Space Agency,
ASI contract ARS--96--70.

\section{REFERENCES}
\vspace{-5mm}
\begin{itemize}
\setlength{\itemindent}{-8mm}
\setlength{\itemsep}{-1mm}

\item[]
Akiyama, M. et al. Proceedings of the XMM meeting (1998) (astro-ph/9811012).

\item[]
Antonucci, R. \ANNREV, {\bf 31}, 473
(1993)

\item[]
Boella G. et al. \AAS, {\bf 122}, 299 (1997a).

\item[]
Boella G. et al. \AAS, {\bf 122}, 327 (1997b).

\item[]
Boyle, B.J. et al. \MN, {\bf 296}, 1 (1998)

\item[]
Cagnoni, I., Della Ceca, R. \& Maccacaro, T.  \APJ, {\bf 493}, 54 (1998).

\item[]
Comastri, A. Setti, G., Zamorani, G. \& Hasinger, G. \AAP, {\bf 296}, 1 (1995).

\item[]
Comastri, A. \SAIT, in press (1999) (astro-ph/9809077). 

\item[]
Fabian, A.C., Barcons, X., Almaini, O. \& Iwasawa, K. \MN, {\bf 297}, L11 (1998).
 
\item[]
Fiore F., La Franca F., Giommi P., Elvis M., Matt G., Comastri A., Molendi
 S. \& Gioia I.M. \MN, submitted (1999) 

\item[]
Francis P.J. \APJ, {\bf 405}, 119  (1993) 

\item[]
Gendreau K.C., Barcons X., Fabian A.C.  \MN, {\bf 297}, 41 (1998)  

\item[]
Georgantopoulos, I., et al. \MN, {\bf 291}, 203 (1997).

\item[]
Giommi P., Fiore F., Ricci D., Molendi S., Maccarone M.C., Comastri A.
In ``The Active X-ray Sky: Results from BeppoSAX and
Rossi-XTE", L. Scarsi, H. Bradt, P. Giommi and F. Fiore (eds.),
Elsevier Science B.V., Nuclear Physics B Proceedings Supplements,
69/1--3, 591 (1998) 

\item[]
Hasinger, G. et al. \AAP, {\bf 329}, 482 (1998).

\item[]
Halpern, J.P., Eracleous, M. \& Forster, K.  \APJ,  {\bf 501}, 103 (1998).

\item[]
Madau, P., Ghisellini, G. \& Fabian A.   \MN, {\bf 270}, L17 (1994)

\item[]
Matt G., \& Fabian A.C. \MN, {\bf 267}, 187 (1994) 

\item[]
Miyaji T., Hasinger G., Schmidt M., in Highlights in X--ray Astronomy in Honor
of Joachim Tr\"umper 65$^{th}$ birthday, MPE Report, in press (1999a)
(astro-ph/9809398)

\item[]
Miyaji T., Hasinger G., Schmidt M., \ASR, in press (1999b)

\item[]
Ogasaka Y., Kii T., Ueda Y., et al. \AN,  {\bf 319}, 43 (1998)

\item[]
Otha K., et al., 1996, \APJL, {\bf 458}, L57 (1996)

\item[]
Page M.J., Carrera F.J., Hasinger G., et al.  \MN, {\bf 281}, 579 (1996) 

\item[]
Piccinotti, G. et al. \APJ, {\bf 253}, 485 (1982).

\item[]
Schmidt, M. et al. \AAP, {\bf 329}, 495 (1998).

\item[]
Setti, G. \& Woltjer, L.  \AAL, {\bf 224}, L21 (1989).

\item[]
Ueda, Y. et al. \NAT, {\bf 391}, 866 (1998).

\item[]
Wood, K.S., et al. \APJS, {\bf 56}, 507 (1984).

\end{itemize}

\end{document}